\documentclass[aps,showpacs,preprint,superscriptaddress,prl]{revtex4}

\usepackage{amsmath}
\usepackage{amssymb}
\usepackage{amscd}
\usepackage{wasysym}
\usepackage[ansinew]{inputenc}
\usepackage[T1]{fontenc}
\usepackage{ae,aecompl}

\usepackage[dvips]{graphicx}
\DeclareGraphicsExtensions{.eps}

\newcommand{\be}{\begin{equation}}
\newcommand{\ee}{\end{equation}}
\newcommand{\bea}{\begin{eqnarray}}
\newcommand{\eea}{\end{eqnarray}}
\newcommand{\nn}{\nonumber}
\newcommand{\ket}[1]{|#1\rangle}
\newcommand{\bra}[1]{\left\langle#1\right|}

\renewcommand{\vec}[1]{\mathbf{#1}}

\newcommand{\ri}{{\rm i}}
\newcommand{\tr}{{\rm tr}}

\begin{document}
\bibliographystyle{apsrev}

\title{Stochastic resonance driven by quantum shot noise
in superradiant Raman scattering}

\author{D Witthaut}

\affiliation{Max-Planck-Institute for Dynamics and Self-Organization (MPIDS), D--37077 G\"ottingen, Germany}

\affiliation{QUANTOP, Niels Bohr Institute, University of Copenhagen, DK--2100 Copenhagen, Denmark}

\date{\today }

\begin{abstract}
We discuss the effects of noise on the timing and strength of
superradiant Raman scattering from a small dense sample of atoms. 
We demonstrate a genuine quantum stochastic resonance effect,
where the atomic response is largest for an appropriate quantum 
noise level.
The peak scattering intensity per atom assumes its maximum for a 
specific non-zero value of quantum noise given by the square 
root of the number of atoms.
\end{abstract}


\maketitle


\section{Introduction}

Stochastic resonance (SR) is a strongly surprising, yet very general effect
in nonlinear dynamical systems. Against our naive understanding, the response
of a system to an external driving  can be facilitated if an appropriate amount of 
noise is added. In fact, the maximum of the response -- the stochastic resonance -- 
is found if the timescale of the noise matches an intrinsic time scale of the system
\cite{Benz81}. By now, a stochastic resonance has been shown in a variety of 
classical systems ranging from neurosciences to geophysics. An overview can be
found in the review articles \cite{Wies95,Dykm95,Gamm98}. 

In recent years there has been an increased interest in the emergence 
of stochastic resonance phenomena in open quantum systems 
(see \cite{Well04} for a review). These phenomena are quite diverse,
as noise and dissipation can refer to  different effects in quantum 
systems. 
Temperature induces a stochastic resonance in a bistable
quantum system in a similar fashion as in the classical case: The
response to a weak driving of the two states assumes its maximum
for a finite value of the system's temperature \cite{Lofs94a,Well00}.
Furthermore, temperature can also enhance genuine quantum features 
of an open system, as for example the entanglement in a quantum spin 
chain \cite{Huel07}.
A different stochastic resonance effect can be found in terms of the 
\emph{coupling} to the environment. In this case, the coherent response of a 
quantum system assumes its maximum for a non-zero coupling to the environment.
This was demonstrated for non-interacting quantum systems,
e.g. in NMR experiments \cite{Viol00}, as well as for interacting
many-body systems, as for example for the phase coherence and purity 
of a Bose-Einstein condensate \cite{08stores,09srlong,11leaky1,11leaky2}.

In this article we will demonstrate another, even more subtle effect.
We identify a stochastic resonance in the superradiant Raman 
scattering of a light pulse. Scattering is strongest if the time scale
of the pulse is matched to the cooperative emission rate of the
the sample, which is determined by the single-atom emission
rate as well as the quantum shot noise. Thus we demonstrate 
a true \emph{quantum} stochastic resonance, as the response
of the sample assumes its maximum for a specific value of the
quantum shot noise, that is, the uncertainty of the initial state.

\section{The Dicke Model}

Superradiance is a collective effect in a system of many atoms interacting 
via the light field. The spontaneous emission of a photon by one of the 
atoms stimulates the other atoms such that light is emitted in a short pulse.
The timescale of this pulse crucially depends on the presence of noise 
and dissipation, making superradiance experiments an ideal laboratory
for stochastic effects in interacting many-body quantum systems.
Furthermore, the basic principles of superradiance are generally valid and 
easy to understand, whereas the details of the dynamics are extremely 
involved \cite{Gros82}. 

The essential features of superradiance are described by the seminal
Dicke model introduced in \cite{Dick54}  (see \cite{Garr11} for a recent review).
We consider light scattering from a small, dense sample of atoms, such that 
propagation effects of the light field can be neglected.
Then one can eliminate the electric field and obtains a master equation 
for the emitters alone. Furthermore, the atoms are indistinguishable 
such that the dynamics can be described by \emph{collective} variables.
In particular, we define the collective atomic operators
\bea
 &&  \hat J_z = \frac{1}{2N} \sum_j \ket{e}_j \bra{e}_j  - \ket{g}_j \bra{g}_j,  \nn \\
  && \hat J_+ = \frac{1}{N} \sum_j \ket{e}_j \bra{g}_j \, \qquad \hat J_- = \hat J_+^\dagger,
\eea
where $\ket{g}_j$ and $\ket{e}_j$ denote ground and excited state of
the $j$th atom, respectively, and $N$ is the number of atoms in the sample.
Then, $\hat J_+$ creates a symmetric excitation in the atomic sample and 
$\hat J_z$ denotes the population imbalance between ground and excited 
state.  The collective operators satisfy the commutation relations
\be
   [\hat J_z, \hat J_\pm] = \pm \hat J_\pm, \qquad
   [\hat J_+, \hat J_-] = 2 \hat J_z.
\ee
For illustrative purposes it  is often more suitable to work with the hermitian
operators 
\be
   \hat J_x = \frac{1}{2} (\hat J_+ + \hat J_-) 
   \quad  \mbox{and} \quad
   \hat J_y = \frac{1}{2 \ri} (\hat J_+ - \hat J_-) .
\ee
The operators $\hat J_{x,y,z}$ form an angular momentum algebra with
quantum number $j=N/2$. This representation is referred to as the
Bloch representation.

In terms of the collective atomic operators, the Dicke master equation 
for the dynamics of the atoms reads
\be
  \frac{d \hat \rho}{dt} = - \ri [\hat H, \hat \rho] - \frac{\gamma}{2}  
     \left( \hat J_+ \hat J_- \hat \rho + \hat \rho \hat J_+ \hat J_- 
    - 2 \hat J_- \hat \rho \hat J_+ \right).
   \label{eqn:master1}
\ee
and the intensity of the emitted light is given by
\be 
   I(t) = \gamma \, \tr[\hat J_+ \hat J_- \hat \rho(t)].
   \label{eqn:int1}
\ee
The decay rate $\gamma$ is a product of the spontaneous 
emission rate of a single atom and a geometrical factor.

\begin{figure}[tb]
\centering
\includegraphics[width=10cm, angle=0]{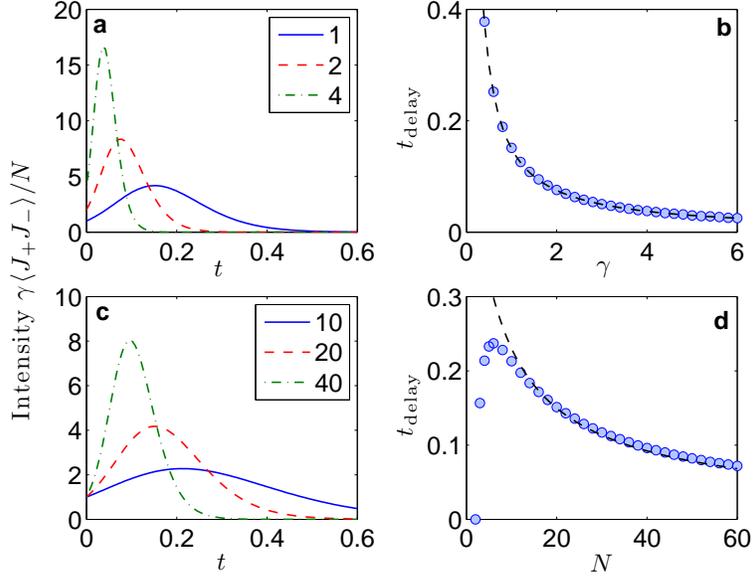}
\caption{\label{fig:delaytime}
Delay of the superradiant pulse as a function of the decay
rate $\gamma$ (upper panels) and the atom number $N$
(lower panels).
(a) Intensity $I(t)$ of the light pulse
for $N = 20$ atoms and three different values of $\gamma$.
(b) Delay time $t_{\rm delay}$ as a function of $\gamma$
for $N=20$ atoms.
(c) Intensity $I(t)$ of the light pulse
for $\gamma = 1$ and three different values of $N$.
(d) Delay time $t_{\rm delay}$ as a function of the atom number
$N$ and $\gamma = 1$.
The dashed line is the mean-field prediction (\ref{eqn:delay-mf}).
}
\end{figure}

As the light intensity (\ref{eqn:int1}) is the essential observable 
measured in experiments, we briefly review some properties 
of the emitted superradiant pulse, in particular the time scale of 
the emission. Assuming that initially
all atoms are in the excited state, each atom can spontaneously emit a
photon. However, the emitted photons also stimulate the emission of further 
photons from the remaining excited atoms. Thus, the number of photons rapidly 
increases such that light is emitted in short intense pulse, as shown in
Fig.~\ref{fig:delaytime} (a) and (c).
The delay time $t_{\rm delay}$, at which the pulse assumes its maximum, obviously 
depends on the emission rate $\gamma$ as shown in Fig.~\ref{fig:delaytime} (b).

Furthermore, the dynamics strongly depends on the quantum noise of the 
initial quantum state. When all atoms are in the excited state, the quantum 
uncertainty of the collective atomic variables are given by
\be
  \Delta J_x = \Delta J_y = \sqrt{N}.
  \label{eqn:qfluct}
\ee
The quantum shot noise, i.e. the \emph{relative} uncertainty of the
Bloch vector $\vec J$, thus scales as
\be
   \frac{|\Delta \vec J|}{|\vec J|} \sim \frac{1}{\sqrt{N}}
\ee
and vanishes in the classical limit $N \rightarrow  \infty$.
This noise drives the emission process such that the
delay time of the superradiant pulse depends strongly on the atom
number $N$. As shown in Fig.~\ref{fig:delaytime} (d), the
delay time first increases rapidly with $N$ and 
then decreases again.  
The delay time can be inferred from a mean-field treatment of the
atomic dynamics, taking into account the quantum fluctuations of
the initial state (\ref{eqn:qfluct}). 
This approximation predicts that \cite{Gros82}
\be
  t_{\rm delay} = (\gamma N)^{-1} \log N.
  \label{eqn:delay-mf}
\ee
The mean-field prediction is compared to the numerically exact
results in Fig.~\ref{fig:delaytime} (b) and (d), showing 
a good agreement when the atom number is large enough.

\section{Driven Superradiant Light Scattering}

\begin{figure}[tb]
\centering
\includegraphics[width=10cm, angle=0]{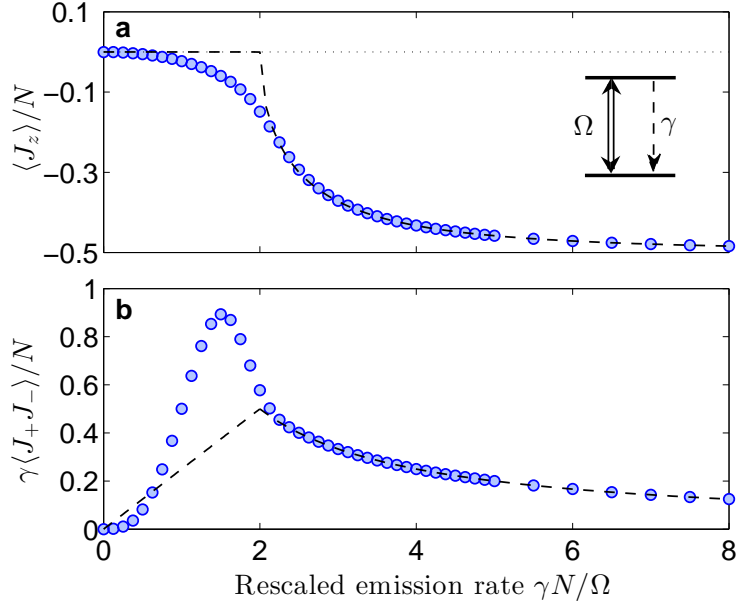}
\caption{\label{fig:steady}
The steady state of the resonantly driven Dicke model.
(a) Population imbalance $\langle \hat J_z \rangle/N$
and (b) intensity of the emitted light (\ref{eqn:int1})
as a function of the rescaled emission rate $\gamma N /\Omega$.
Circles represent numerical results for $N=20$ atoms and $\Omega=1$,
the dashed line is the mean-field prediction (\ref{eqn:mfsteady}).
The inset in (a) illustrates the system under consideration.
}
\end{figure}

The Dicke model shows a very rich dynamics when an external driving by
a classical light field is included.
The Hamiltonian describing the coherent driving is given by
\be
  \hat H = - \Delta \hat J_z - \frac{\Omega}{2} \left( \hat J_+ + \hat J_- \right),
\ee
where $\Omega$ is the Rabi frequency and $\Delta$ is the detuning of
the laser frequency with respect to the atomic transition frequency.
If the driving is resonant ($\Delta = 0$), the Dicke model undergoes 
a quantum phase transition in the thermodynamic limit $N \rightarrow \infty$. 
In this limit we can invoke a mean-field approximation, such that the 
dynamics of the expectation values $J_{z} = \langle \hat J_z \rangle$
and $J_{\pm} = \langle \hat J_\pm \rangle$ follows the equations of motion
\bea
 \frac{d J_+}{dt} &=& -\ri \Delta J_+ + \ri \Omega J_z + \gamma  J_+  J_z \nn \\
 \frac{d J_z}{dt} &=& \ri \frac{\Omega}{2}(J_+ - J_-) - \gamma  J_+  J_- ,
 \label{eqn-pendulum2}
\eea
On the resonance, the steady state is given by
\bea
   J_z &=& \left\{ \begin{array}{l c l}
       0    							 & \mbox{ for } &  2 \Omega  \le  \gamma N \\
       - \sqrt{N^2/4 - \Omega^2/\gamma}    &                     &  2 \Omega  >  \gamma N
   \end{array} \right. \nn \\
   J_+ J_- &=& \left\{ \begin{array}{l c l}
       N^2/4   					   & \mbox{ for } &  2 \Omega  \le  \gamma N \\
       \Omega^2/\gamma^2   &                     &  2 \Omega  >  \gamma N
   \end{array} \right.  
   \label{eqn:mfsteady}
\eea
For strong driving or a weak decay, respectively, the population
imbalance vanishes exactly. When the decay becomes stronger,
$\gamma N > 2 \Omega$, a phase transition to an ordered state with
$\langle \hat J_z \rangle < 0$ takes place. This behavior is shown in
Fig.~\ref{fig:steady} (a), where the population imbalance 
$\langle \hat J_z \rangle$ is plotted as a function of the rescaled 
emission rate $\gamma N/\Omega$. For finite particle number $N$,
the transition becomes smeared out and $\langle \hat J_z \rangle$
is small but non-zero, even for $\gamma N < 2 \Omega$.
The Dicke phase transition can be analyzed fully analytically for
arbitrary particle numbers $N$ \cite{Garr11}, so we do not go into
detail here. An experimental observation of this phase transition
has been reported only recently \cite{Baum10}.

\begin{figure}[tb]
\centering
\includegraphics[width=9cm, angle=0]{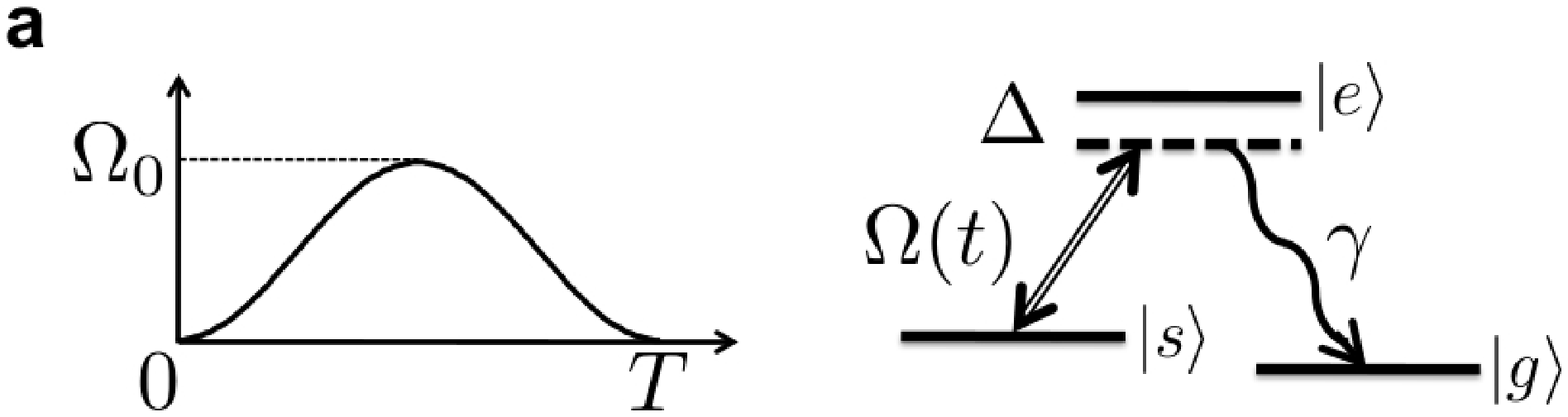}
\includegraphics[width=9cm, angle=0]{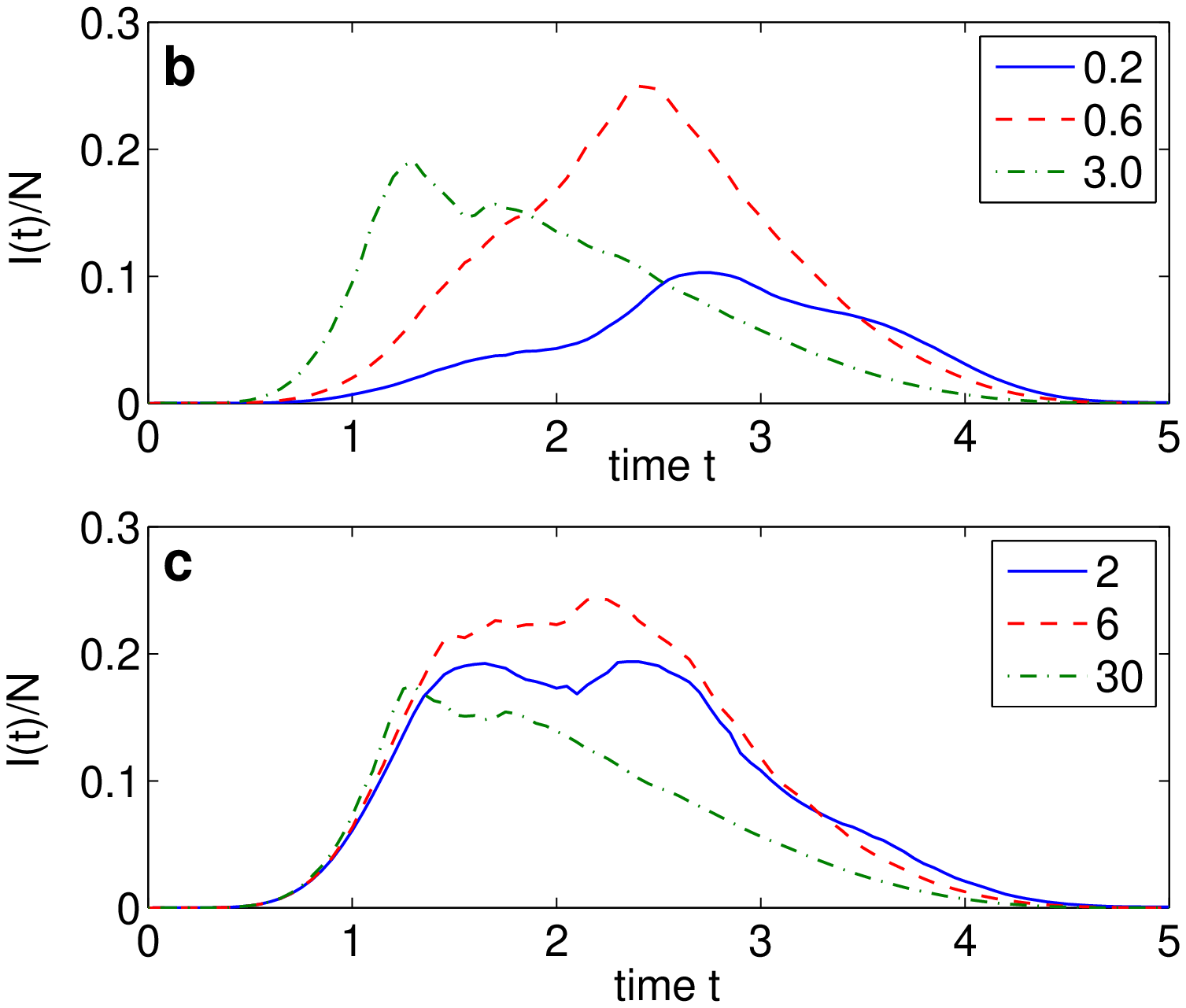}
\caption{\label{fig:srpulsed1}
Superradiant Raman scattering by a sample of $N$ 
$\Lambda$-type atoms.
(a) Schematic representation of the scattering process.
The driving field $\Omega(t)$ is time-dependent with a pulse 
length $T$ and maximum value $\Omega_0$.
(b) The emitted superradiant pulse $I(t)$ for three different values of
$\gamma$ and $N = 20$.
(c) The emitted superradiant pulse $I(t)$ for three different values of
$N$ and $\gamma = 1$.
}
\end{figure}

In this article, we are primarily interested in the scattering light
intensity, which can be viewed as the response of the atomic cloud 
to the coherent driving. 
Figure \ref{fig:steady} (b) shows the intensity (\ref{eqn:int1}) per
atom as a function of the emission rate $\gamma$. One observes
a clear maximum in the vicinity of the phase transition 
$\gamma N \approx 2 \Omega$. 
This can be viewed as a stochastic resonance effect. The
response of the quantum system assumes its maximum 
for a specific non-zero coupling to the environment, given
by a specific value of $\gamma$. Moreover, the maximum 
is found if the timescale of the driving $(2 \Omega)^{-1}$
is matched to the timescale of the environment coupling
$(N \gamma)^{-1}$.

We note that the scattered intensity essentially depends 
on the effective emission rate $\gamma N/\Omega$.
Hence, the observed stochastic resonance is also present 
in the scattering of a single atom and when the atom 
number $N$ is varied.
This effect is similar to the stochastic resonance observed
in systems described by the Bloch equations, which 
are also governed by the competition of excitation 
and relaxation \cite{Viol00,08stores,09srlong}.
Stochastic resonance effects in composite classical systems
and their dependence on the number of constituents has
been analyzed in \cite{Marc96}.

\section{Stochastic Resonance in Periodically Driven Systems}

\begin{figure}[tb]
\centering
\includegraphics[width=8cm, angle=0]{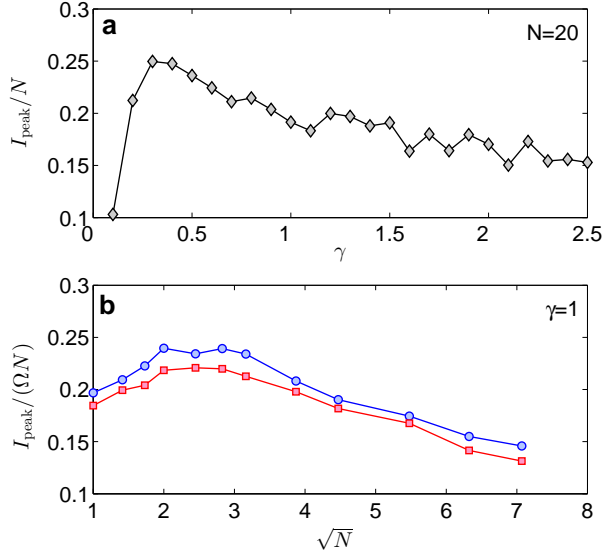}
\caption{\label{fig:srpulsed2}
Stochastic resonance in superradiant Raman scattering: The
peak intensity per atom of the emitted pulse assumes a maximum
for a specific non-zero value of
(a) the coupling rate to the environment $\gamma$ and
(b) the magnitude of the quantum uncertainty $\sqrt{N}$.
The remaining parameters are $\Omega_0 = 1$ (blue circles) and
$\Omega_0 = 2$ (red squares), $\Delta = 1$, 
$T = 5$ and (a) $N=20$ and (b) $\gamma=1$, respectively.
Solid lines are shown to guide the eye.
}
\end{figure}

Superradiant light emission is essentially started by quantum 
fluctuations such that its timescale is determined by the
quantum uncertainty $N^{1/2}$. Therefore we can identify 
another, much more subtle stochastic resonance effect.

We consider superradiant Raman scattering of a light pulse by
a dense sample of three-level atoms as shown in Fig.~\ref{fig:srpulsed1} (a).
The atoms emit light collectively, when they decay from the excited
state $\ket{e}$ to the ground state $\ket{g}$ as described by
the Dicke master equation (\ref{eqn:master1}). However, we
assume that they are initially prepared in a metastable state 
$\ket{s}$, which is coupled to the excited state by a classical light pulse,
\be
   \hat H = - \frac{\Delta}{2}  \left( \hat J_{ss} + \hat J_{ee} \right)
                - \frac{\Omega(t)}{2} \left( \hat J_{se} + \hat J_{es} \right).
\ee
The collective operators used here are defined as usual
\be
  \hat J_{ab} = \frac{1}{N} \sum_j \ket{a}_j \bra{b}_j \, ,
\ee
where $a$ and $b$ label one of the atomic states $\ket{s}$ or $\ket{e}$.
For simplicity, we assume that the input pulse is given by
$\Omega(t) = \Omega_0 \sin(\pi t/T)^2$ in the interval
$t \in [0,T]$. We solve the master equation numerically using
the Monte Carlo wave function method \cite{Dali92,Carm93}
averaging over 500 ($N \le 15$) resp. 100 ($N \ge 20$) random trajectories.

Figure \ref{fig:srpulsed1} (a) and (b) show the emitted light pulse
for different values of $\gamma$ and the atom number $N$.
In both cases, we observe that the peak intensity is largest for
intermediate values of $\gamma$ or $N$, respectively.
Obviously, the peak intensity is rather small if the emission rate
$\gamma$ and the atom number $N$ are small. In this case the 
emission is too weak and most atoms remain in the initial state $\ket{s}$.
However, if $\gamma$ or $N$ are too large, then the emission
starts 'too early', long before the driving $\Omega(t)$ has reached
its maximum. Also in this case, the peak intensity is limited.

This is further analyzed in Fig.~\ref{fig:srpulsed2}, where the peak 
intensity per atom is plotted as a function of (a) the emission 
rate $\gamma$ and $(b)$ the atom number $N$.
In both cases we observe a clear maximum when $\gamma$ or 
$N$ assume intermediate values, such that the time scale of the 
superradiant emission matches the timescale of the driving. 
The second case is particularly interesting, as the maximum
is found for a specific non-zero strength of the quantum uncertainty
$N^{1/2}$.
The observed stochastic resonance is thus a \emph{genuine quantum}
effect, as it is observed in terms of the quantum shot noise. 

\section{Conclusion}

In this article
we have analyzed different stochastic resonance effects in 
superradiant light scattering. We have identified a genuine quantum 
stochastic resonance, where the response of the atomic sample
to an external driving assumes its maximum for a specific 
non-zero strength of the quantum shot noise $N^{-1/2}$.
The response is quantified by the scattered light intensity 
per atom or its peak value, respectively.

The mechanism generating the stochastic resonance is a 
matching of the timescales of the driving and the superradiant 
emission.  
The latter depends on the emission rate $\gamma$, but also on the quantum 
shot noise of the sample, as the the emission process is essentially 
started by quantum fluctuations. Thus we can observe a maximum of
the response, the stochastic resonance, in terms of both the coupling
rate to the environment $\gamma$
and the quantum uncertainty $N^{1/2}$.

\section*{Acknowledgements}

I thank A Hilliard and J H M\"uller for stimulating discussions. 
Financial support from the Deutsche Forschungsgemeinschaft 
(grant number DFG WI 3415/1-1) and the Max Planck society 
is gratefully acknowledged.



\begin{thebibliography}{10}


\bibitem{Benz81} 
Benzi R, Sutera A and Vulpiani A (1981) 
\textit{J. Phys. A: Math. Gen.} {\bf 14} L453

\bibitem{Wies95} 
Wiesenfeld K and Moss F (1995) 
\textit{Nature (London)} {\bf 373} 33

\bibitem{Dykm95}
Dykman M I, Luchinsky D G, Mannella R, McClintock P V E, 
Stein N D and Stocks N G (1995) 
\textit{Nuovo Cimento D} {\bf 17} 661

\bibitem{Gamm98} 
Gammaitoni L, H\"anggi P, Jung P and Marchesoni F (1998)
\textit{Rev. Mod. Phys.} {\bf 70} 223

\bibitem{Well04}
Wellens T, Shatokhin V and Buchleitner A (2004)
\textit{Rep. Prog. Phys.} {\bf 67} 45

\bibitem{Lofs94a}
L\"{o}fstedt R and Coppersmith S N (1994)
\textit{Phys. Rev. Lett.} {\bf 72} 1947

\bibitem{Well00}
Wellens T  and Buchleitner A (2000)
\textit{Phys. Rev. Lett.} \textbf{84} 5118

\bibitem{Huel07}
Huelga S F and Plenio M B (2007) 
\textit{Phys. Rev. Lett.} {\bf 98} 170601

\bibitem{Viol00}
Viola L, Fortunato E M, Lloyd S, Tseng C H and Cory D G (2000)
\textit{Phys. Rev. Lett.} {\bf 84} 5466 

    
\bibitem{08stores}
Witthaut D, Trimborn F and Wimberger S (2008)
\textit{Phys.~Rev.~Lett.} {\bf 101} 200402

\bibitem{09srlong}
Witthaut D, Trimborn F and Wimberger S (2009)
\textit{Phys.~Rev.~A} \textbf{79} 033621

\bibitem{11leaky1}
Witthaut D, Trimborn F, Hennig H, Kordas G, Geisel T and Wimberger S (2011)
\textit{Phys.~Rev.~A} \textbf{83} 063608

\bibitem{11leaky2}
Trimborn F, Witthaut D, Hennig H, Kordas G, Geisel T and Wimberger S (2011)
\textit{Eur.~Phys.~J.~D} \textbf{63} 63



\bibitem{Gros82}
Gross M and Haroche S (1982)
\textit{Phys. Rep.} {\bf 93} 301
(1992) 
\textit{Phys. Rev. Lett.} {\bf 68} 580
\bibitem{Dick54}
Dicke R H (1954)
\textit{Phys. Rev.} \textbf{93} 99 

\bibitem{Garr11}
Garraway B M (2011)
\textit{Phil. Trans. R. Soc. A} \textbf{369} 1137


\bibitem{Baum10}
Baumann K, Guerlin C, Brennecke F  and Esslinger T (2010) 
\textit{Nature (London)} \textbf{464} 1301


\bibitem{Marc96}
Marchesoni F, Gammaitoni L and Bulsara A R (1996) 
\textit{Phys. Rev. Lett.} {\bf 76} 2609


\bibitem{Dali92}
Dalibard J, Castin Y and M\o{}lmer K (1992) 
\textit{Phys. Rev. Lett.} {\bf 68} 580

\bibitem{Carm93}
Carmichael H J  (1993)
\textit{An Open Systems Approach to Quantum Optics} 
(Springer, Berlin)

\end{thebibliography}
\end{document}